# Robust Steganography Using LSB-XOR and Image Sharing

Chandranath Adak
Department of Computer Science and Engineering
University of Kalyani
Kalyani – 741235 , India
adak32@gmail.com

*Abstract*—Hiding and securing the secret digital information and data that are transmitted over the internet is of widespread and most challenging interest. This paper presents a new idea of robust steganography using bitwise-XOR operation between stego-key-image-pixel LSB (Least Significant Bit) value and secret message- character ASCII-binary value (or, secret image-pixel value). The stego-key-image is shared in dual-layer using odd-even position of each pixel to make the system robust. Due to image sharing, the detection can only be done with all the image shares.

## I. INTRODUCTION

Steganography has gained significance over years for its art of writing hidden messages in such a way, that no one, apart from the sender and intended recipient, suspects the existence of the message, it is a form of security through obscurity[1]. The term *steganography* is derived from two Greek words *steganos*, meaning "covered", and *graphein*, meaning "to write". The first recorded use of the term was in 1499 by Johannes Trithemius in his *Steganographia*, a treatise on cryptography and steganography disguised as a book on magic[2]. The history of steganography can be traced back to around 440 B.C., where the Greek historian Herodotus described in his writing about two events : one used wax to cover secret messages, and the other used shaved heads[3]. Modern steganography entered the world in 1985 with two engineers, Barrie Morgan and Mike Barney[4].Digital image is the most popular carrier in the study of steganography. Steganography differs from cryptography in the sense that where cryptography focuses on keeping the contents of a message secret, steganography focuses on keeping the existence of a message secret[5-11].

There has been many approaches in steganography, which includes StegHide[12], OutGuess[13], model based steganography[14], perturbed quantization[15] and statistical restoration[16], but the most traditional approach is hiding data in LSB[17] of spatial or transform domain coefficient.

Naor & Shamir[18] demonstrated a visual secret sharing scheme, where an image was broken up into *n* shares so that only someone with all *n* shares could decrypt the image, while any *n-1* shares revealed no information about the original image. In the prescribed cryptosystem, the stego-key-image is broken into two slides.

Here, we are presenting a stego-system using digital image as a carrier (Stego Key), and LSB-XOR technique with image sharing concept of visual cryptography.

## II. STEPS OF PROPOSED METHOD

### A. Stego-Key Generation

An arbitrary grey image (Stego-Key-image) is chosen for stego-key generation. The grey value of each pixel is extracted from this image and using odd-even position of those pixels two shared image (S_Key img1 and S_Key img2) is created. The grey values of these shared images are converted into equivalent eight-bit binary numbers and the pool of these LSBs is the stego-key values.

### B. Embedding Algorithm

The secret message is either 'textual message' or 'image message' ; for textual-message the ASCII (American Standard Code for Information Interchange) value of each character and for image-message the grey value of each pixel is used. The ASCII values (or, grey values) are treated as decimal numbers and converted into equivalent eight-bit binary numbers.

These eight-bit binary values are stored by its odd-even position to make S_Msg1 (using only odd positioned four bits) and S_Msg2 (using only even positioned four bits).

The S_Msg1 and S_Msg2's bits are bitwise-XORed with LSBs of S_Key img1 and S_Key img2 respectively. After this bitwise XOR operation, the new binary grey values are converted into equivalent decimal grey values. These grey values are stored in odd-even position to form Stego-Image. This Stego-Image is transmitted through the channel.

### C. Detector Algorithm

The Stego-Image's pixel values are treated as decimal numbers and converted into eight bit binary numbers. The Stego-Image is shared in two parts (S_Img1 and S_Img2) with odd-even positioned pixel grey values.

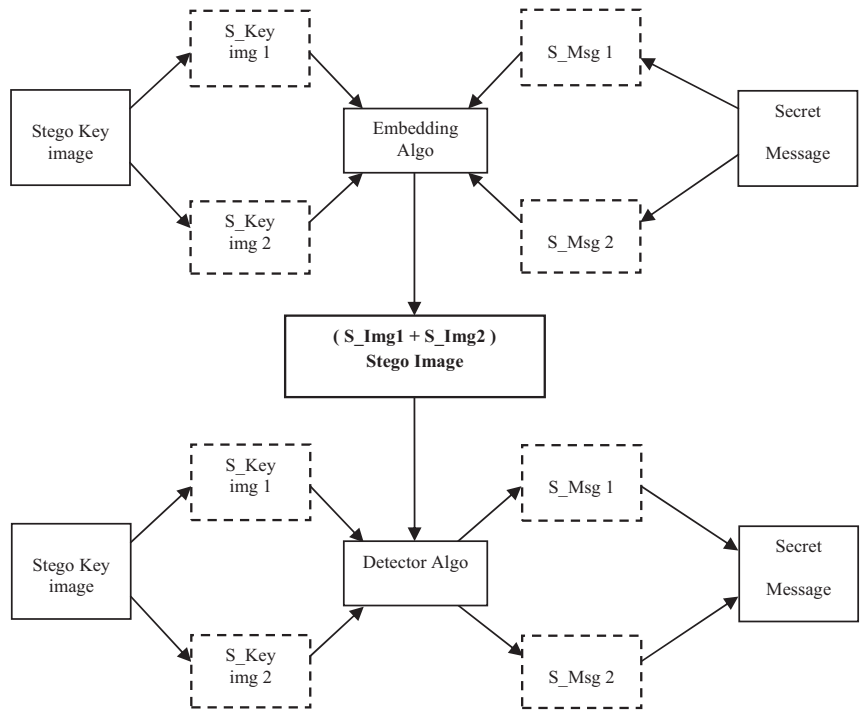

*fig. 1 :* Model of the proposed method

The S_Img1 and S_Img2's LSBs are bitwise XORed with LSBs of S_Key img1 and S_Key img2 respectively.

The bitwise-XOR operated LSBs are stored by odd-even position and taken as pool of eight bit binary numbers. These eight-bit binary numbers are converted into equivalent decimal numbers and seemed as either ASCII values for textual-message or grey values for image-message. The secret message is detected with these values.

The model of this proposed method is shown in *fig.1* .

### III. IMPLEMENTATION

The following is the illustration of the proposed methodology with an example.

Let, the Secret Message is textual message and it contains a single letter "I". The secret Stego-Key-image is the Lena image.

In the embedding process, at sender side, the odd and even positioned Stego-Key-image pixels form S_Key img1 and S_Key-img2 respectively by image sharing in dual part as shown in *table 1*.

| Pixel Co-ordinate of Stego-Key Image | Stego-Key-image-pixel grey values | Equivalent 8-bit binary |
|---|---|---|
| (0,0) | 162 | 1010001**0** |
| (0,1) | 161 | 1010000**1** |
| (0,2) | 158 | 1001111**0** |
| (0,3) | 156 | 1001110**0** |
| (0,4) | 156 | 1001110**0** |
| (0,5) | 153 | 1001100**1** |
| (0,6) | 154 | 1001101**0** |
| (0,7) | 161 | 1010000**1** |
| (0,8) | 168 | 1010100**0** |
| (0,9) | 173 | 1010110**1** |
| . | . | . |
| . | . | . |
| . | . | . |

Table 1.(a)

| Pixel Co-ordinate of S_Key img 1 | Pixel Co-ordinate of Stego-Key-image | S_Key img 1-pixel grey values | Equivalent 8-bit binary (B_SKey1) |
|---|---|---|---|
| (0,0) | (0,0) | 162 | 1010001**0** |
| (0,1) | (0,2) | 158 | 1001111**0** |
| (0,2) | (0,4) | 156 | 1001110**0** |
| (0,3) | (0,6) | 154 | 1001101**0** |
| (0,4) | (0,8) | 168 | 1010100**0** |
| . | . | . | . |
| . | . | . | . |
| . | . | . | . |

Table 1.(b)

| Pixel Co-ordinate of S_Key img 2 | Pixel Co-ordinate of Stego-Key-image | S_Key img 2-pixel grey values | Equivalent 8-bit binary (B_SKey2) |
|---|---|---|---|
| (0,0) | (0,1) | 161 | 1010000**1** |
| (0,1) | (0,3) | 156 | 1001110**0** |
| (0,2) | (0,5) | 153 | 1001100**1** |
| (0,3) | (0,7) | 161 | 1010000**1** |
| (0,4) | (0,9) | 173 | 1010110**1** |
| . | . | . | . |
| . | . | . | . |
| . | . | . | . |

Table 1.(c)

*Table 1.(a),(b),(c)* show Stego-Key-image, S_Key-img1 and S_Key-img2 pixel grey values and their equivalent 8-bit binary. The LSBs are shown in bold format.

The Secret Message (textual) is "I", its ASCII value is '73', which is treated as decimal number and converted into eight bit binary number. $(73)_{10} = (01001001)_2$. This binary number is stored in the reverse way and shared in two parts with respect to odd-even position as shown in *fig.2*.

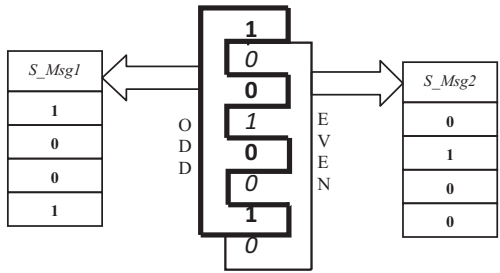

***fig. 2 :*** Secret-Message partitioning

B_SKey1 and B_SKey2 are the equivalent eight-bit binary of S_Key-img1 and S_Key-img2 pixel grey values respectively. The LSBs of B_SKey1 and B_SKey2 are bitwise-XORed with S_Msg1 and S_Msg2 respectively as *table 2*.

| LSB of B_SKey1 (l1) | S_Msg1 (sm1) | i1 = l1 ^ sm1 |
|---|---|---|
| 0 | 1 | 1 |
| 0 | 0 | 0 |
| 0 | 0 | 0 |
| 0 | 1 | 1 |

Table 2.(a)

| LSB of B_SKey2 (l2) | S_Msg2 (sm2) | i2 = l2 ^ sm2 |
|---|---|---|
| 1 | 0 | 1 |
| 0 | 1 | 1 |
| 1 | 0 | 1 |
| 1 | 0 | 1 |

Table 2.(b)

*Table 2.(a)* shows the bitwise XOR operation between LSB of B_SKey1 and S_Msg1 ; *Table 2.(b)* shows the bitwise XOR operation between LSB of B_SKey2 and S_Msg2 .

The i1 and i2 is replaced in the LSBs of B_SKey1 and B_SKey2 to form S_Img1 and S_Img2's pixel binary grey values respectively as *table 3.(a),(b)* .

| B_SKey1 | i1 | S_Img1 pixel binary grey value (replacing LSB of B_SKey1 with i1) | S_Img1 pixel decimal grey value (after replacement) |
|---|---|---|---|
| 1010001**0** | 1 | 1010001**1** | 163 |
| 1001111**0** | 0 | 1001111**0** | 158 |
| 1001110**0** | 0 | 1001110**0** | 156 |
| 1001101**0** | 1 | 1001101**1** | 155 |

Table 3.(a)

| B_SKey2 | i2 | S_Img2 pixel binary grey value (replacing LSB of B_SKey2 with i2) | S_Img2 pixel decimal grey value (after replacement) |
|---|---|---|---|
| 1010000**1** | 1 | 1010001**1** | 161 |
| 1001110**0** | 1 | 1001111**1** | 157 |
| 1001100**1** | 1 | 1001110**1** | 153 |
| 1010000**1** | 1 | 1001101**1** | 161 |

Table 3.(b)

| Pixel Co-ordinate of Stego-Image | Stego-Image-pixel binary grey values | Equivalent decimal grey values |
|---|---|---|
| (0,0) | 1010001**1** | 163 |
| (0,1) | 1010001**1** | 161 |
| (0,2) | 1001111**0** | 158 |
| (0,3) | 1001111**1** | 157 |
| (0,4) | 1001110**0** | 156 |
| (0,5) | 1001110**1** | 153 |
| (0,6) | 1001101**1** | 155 |
| (0,7) | 1001101**1** | 161 |
| (0,8) | 1010100**0** | 168 |
| (0,9) | 1010110**1** | 173 |
| . | . | . |
| . | . | . |
| . | . | . |

Table 3.(c)

*Table 3.(a),(b),(c)* show the formation of Stego-Image.

The S_Img1 and S_Img2 pixel decimal grey values are stored by odd-even position respectively ( as *table 3.(c)* ) and remaining unaltered pixel grey values ( from pixel co-ordinate (0,8) as *table 3.(c)* ) of Stego-Key-Image are simply copied to form Stego-Image.

Sender transmits this Stego-Image to the receiver. The receiver uses the detector algorithm to extract the secret message from the stego image.

In the detector process, at the receiver side, the Stego-Image is shared in two parts with respect to odd-even position of its pixel values as *table 4*.

| Pixel Co-ordinate of Stego-Image | Stego-Image-pixel grey values | Equivalent 8-bit binary |
|---|---|---|
| (0,0) | 163 | 1010001**1** |
| (0,1) | 161 | 1010001**1** |
| (0,2) | 158 | 1001111**0** |
| (0,3) | 157 | 1001111**1** |
| (0,4) | 156 | 1001110**0** |
| (0,5) | 153 | 1001110**1** |
| (0,6) | 155 | 1001101**1** |
| (0,7) | 161 | 1001101**1** |
| (0,8) | 168 | 1010100**0** |
| (0,9) | 173 | 1010110**1** |
| . | . | . |

Table 4.(a)

| Pixel Co-ordinate of S_Img1 | Pixel Co-ordinate of Stego-Image | S_Img1-pixel grey values | Equivalent 8-bit binary (B_SImg1) |
|---|---|---|---|
| (0,0) | (0,0) | 163 | 1010001**1** |
| (0,1) | (0,2) | 158 | 1001111**0** |
| (0,2) | (0,4) | 156 | 1001110**0** |
| (0,3) | (0,6) | 155 | 1001101**1** |
| (0,4) | (0,8) | 168 | 1010100**0** |
| . | . | . | . |

Table 4.(b)

| Pixel Co-ordinate of S_Img2 | Pixel Co-ordinate of Stego-Image | S_Img2-pixel grey values | Equivalent 8-bit binary (B_SImg2) |
|---|---|---|---|
| (0,0) | (0,1) | 161 | 1010001**1** |
| (0,1) | (0,3) | 157 | 1001111**1** |
| (0,2) | (0,5) | 153 | 1001110**1** |
| (0,3) | (0,7) | 161 | 1001101**1** |
| (0,4) | (0,9) | 173 | 1010110**1** |
| . | . | . | . |

Table 4.(c)

*Table 4.(a),(b),(c)* show the Stego_Image sharing.

B_SImg1 and B_SImg2 are the equivalent eight-bit binary of S_Img1 and S_Img2 pixel grey values respectively.

The LSBs of B_SKey1 and B_SKey2 are bitwise-XORed with LSBs of B_SImg1 and B_SImg2 respectively as table 5.

| LSB of B_SKey1 (l1) | S_Img1 (si1) | m1 = l1 ^ si1 |
|---|---|---|
| 0 | 1 | 1 |
| 0 | 0 | 0 |
| 0 | 0 | 0 |
| 0 | 1 | 1 |

Table 5.(a)

| LSB of B_SKey2 (l2) | S_Img1 (si2) | m2 = l2 ^ si2 |
|---|---|---|
| 1 | 1 | 0 |
| 0 | 1 | 1 |
| 1 | 1 | 0 |
| 1 | 1 | 0 |

Table 5.(b)

*Table 5.(a)* shows the bitwise XOR operation between LSB of B_SKey1 and S_Img1 ; *Table 5.(b)* shows the bitwise XOR operation between LSB of B_SKey2 and S_Img2 .

The m1 and m2 are basically S_Msg1 and S_Msg2 respectively. The S_msg1 bit values are placed in odd position and S_Msg2 bit values are placed in even position to form a sequence of zeros and ones as *fig. 3*.

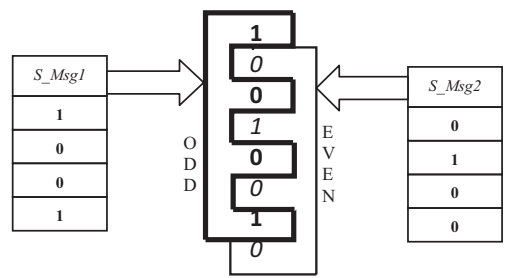

*fig. 3 :* Secret-Message formation

This sequence is sliced in eight bit block; each block is treated as an eight bit binary number. This binary number is stored in reverse way as '01001001' and converted into its equivalent decimal. $(01001001)_2 = (73)_{10}$. This decimal number is treated as ASCII value, whose equivalent character is 'I'. And secret textual message was "I". So, the Secret_Message is detected with the help of this prescribed method.

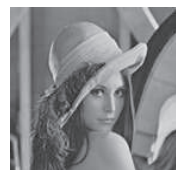
*fig. 4.(a1)*

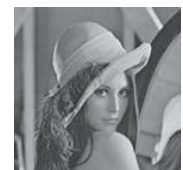
*fig. 4.(a2)*

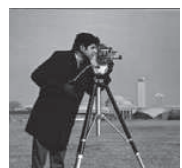
*fig. 4.(b1)*

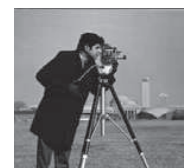
*fig. 4.(b2)*

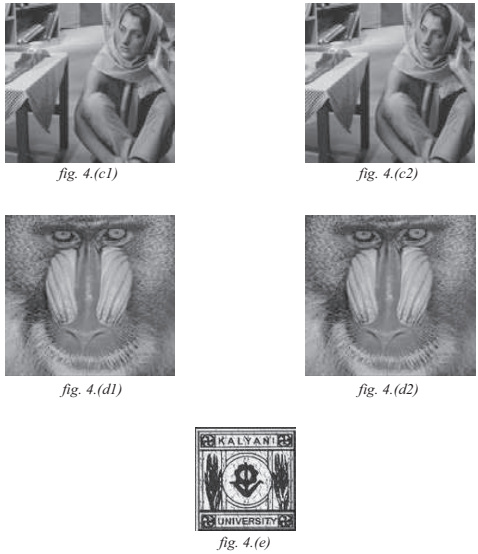

fig. 4.(c1)    fig. 4.(c2)

fig. 4.(d1)    fig. 4.(d2)

fig. 4.(e)

fig. 4.(a1),(b1),(c1),(d1) are the Stego_Key image.
fig.4.(a2),(b2),(c2),(d2) are the Stego_Image with embedded Secret_Message "I" , "Computer" , "Department of Computer Science and Engineering  - University of Kalyani" and the logo of fig.4.(e). All the Stego_Key images and Stego_Images are of dimension 100X100.

The Stego_Key image and Stego_Image can be compared using histogram analysis.

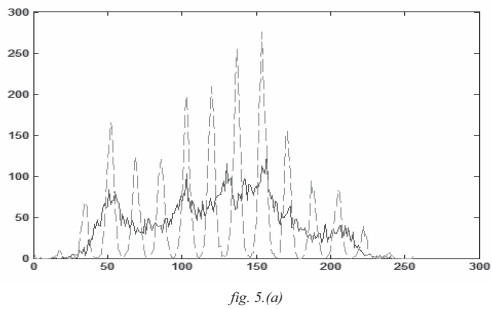

fig. 5.(a)

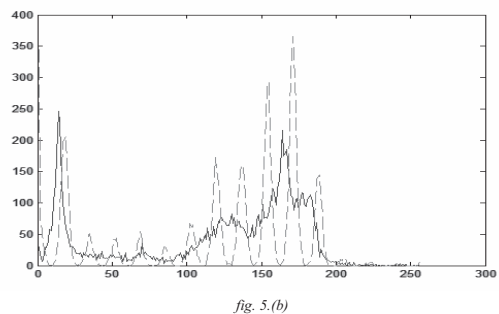

fig. 5.(b)

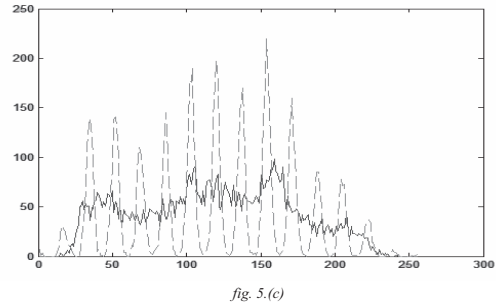

fig. 5.(c)

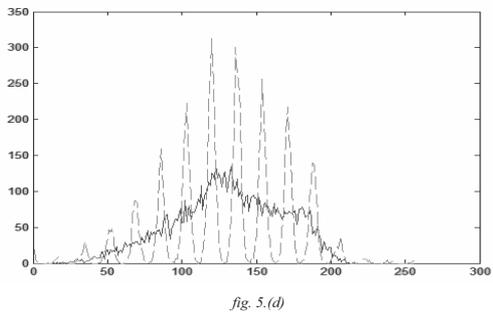

fig. 5.(d)

fig. 5.(a),(b),(c),(d) show the histogram comparison between Stego_Key image ( fig. 4.(a1),(b1),(c1),(d1) – shown by continuous line ) and Stego_Image ( fig. 4.(a2),(b2),(c2),(d2) – shown by discontinuous line ) .

From this histogram analysis of the Stego_Key image and Stego_Image, it is cleared that the secret messages are well embedded in the Stego_Key image.

### IV.    CONCLUSION AND FUTURE WORK

The proposed methodology has been tested for huge amount of different textual and image secret messages, and it works satisfactory; it has high robustness, advanced level security and powerful embedding system. The secret message is hiding behind the stego-key-image, so the existence of the message is more secure. The bitwise XOR operation is done with only the LSBs of binary grey values of the stego-key-image, so the stego image pixel either remains same or increases (or, decreases) by one. The stego-key image is shared in two parts, so without getting proper image-shares the secret message cannot be detected. The next venture of this prescribed methodology is to make this genetic algorithm based steganographic system [19-20].

ACKNOWLEDGMENT

I would like to heartily thank Prof. Bidyut B. Chaudhuri, Head, Computer Vision and Pattern Recognition Unit, Indian